\documentclass[12pt]{article}

\usepackage{latexsym}

\usepackage{graphicx}

\begin{document}


\title{Magnetized configurations with black holes and  Kaluza-Klein bubbles: Smarr-like relations \\  and first law}

\author{
     Stoytcho S. Yazadjiev \thanks{E-mail: yazad@phys.uni-sofia.bg}, Petia G. Nedkova\thanks{E-mail:pnedkova@phys.uni-sofia.bg}\\
{\footnotesize  Department of Theoretical Physics,
                Faculty of Physics, Sofia University,}\\
{\footnotesize  5 James Bourchier Boulevard, Sofia~1164, Bulgaria }\\
}

\date{}

\maketitle

\begin{abstract}
We present a general class of exact solutions in Einstein-Maxwell-dilaton gravity
describing  configurations of black holes and Kaluza-Klein bubbles  magnetized along the compact dimension.
Smarr-like relations for the mass and the tension are found. We also derive the mass and tension  first law for the
configurations under consideration using the Noether current approach.
The solutions we consider are explicit examples showing  that in Kaluza-Klein spacetimes the
interval (rod) structure and the charges (which are zero  by construction for the solutions here),
are insufficient to classify the solutions and  additional data is necessary, namely the magnetic
flux(es).

\end{abstract}


\sloppy

\section{Introduction}

In recent years the higher dimensional gravity and especially black holes in higher dimensions  have attracted a lot of interest.
During the last decade many exact black solutions were found and investigated from different perspectives \cite{ER1}.
Most of the known exact solutions are for asymptotically
flat spacetimes. We could say that we have rather complete understanding of the asymptotically flat black holes in 5D reflected
in the uniqueness theorems \cite{HY1},\cite{HY2}. Very recent numerical results \cite{KKR} show that some of the properties
of the asymptotically flat 5D black holes might hold also in spacetimes with $D>5$.

While asymptotically flat spacetimes with $D>4$ are an excellent theoretical laboratory giving valuable intuition, higher dimensional
spacetimes with compact extra dimensions (Kaluza-Klein spacetimes\footnote{More precisely, a five dimensional  spacetime is called Kaluza-Klein
spacetime if  it is asymptotically ${\cal M}\times S^1$ where ${\cal M}$ is the Minkowski spacetime.     }) are  more realistic. That is why it is physically important to study black holes in
spacetimes with compact dimensions. Very recently the uniqueness theorem for vacuum Kaluza-Klein black holes was established in \cite{HY3}.
This theorem gives complete  classification of the  possible horizon topologies and classification of the black solutions on the basis of the so-called
interval (rod) structure. Some exact Kaluza-Klein black hole solutions have also been constructed \cite{ER2}-\cite{KY}. Among them are the solutions describing
sequences of static vacuum black holes and bubbles \cite{EH}-\cite{EHO} which are of particular interest in the context of the present paper.
We also refer the reader to the review article \cite{HNO} where the Kaluza-Klein black holes are considered from different perspectives.
Nevertheless, the known exact Kaluza-Klein black hole solutions are far from being exhaustive.
In the present paper we present a general class of exact Kaluza-Klein solutions of 5D Einstein-Maxwell-dilaton equations describing configurations of black holes and Kaluza-Klein bubbles magnetized along the compact dimension.  The basic physical quantities of these solutions are calculated. We also derive  Smarr-like relations  as well as the first law for the mass and the tension. Let us stress that we use "black holes" to denote any black objects independently of their horizon topologies.

The solutions of the present paper are also important as explicit examples showing some subtleties   which should be taken
into account in the generalization of the uniqueness theorem of \cite{HY3} to the case with Maxwell fields.

\section{Exact solutions}

In five dimensions the Einstein-Maxwell-dilaton gravity is described by the field equations
\begin{eqnarray}\label{EMDFE}
&&R_{\mu\nu}= 2\partial_{\mu}\varphi\partial_{\nu}\varphi +2e^{-2\alpha\varphi}\left(F_{\mu\sigma}F_{\nu}^{\,\sigma}
- {1\over 6}g_{\mu\nu}F_{\lambda\sigma}F^{\lambda\sigma}\right), \nonumber \\ \nonumber \\
&&\nabla_{\mu}\left(e^{-2\alpha\varphi}F^{\mu\nu} \right)=0 , \;\; \nabla_{[\sigma} F_{\mu\nu]}=0, \\ \nonumber \\
&&\nabla_{\mu}\nabla^{\mu}\varphi = - {\alpha\over 2}e^{-2\alpha\varphi}F_{\sigma\lambda}F^{\sigma\lambda}, \nonumber
\end{eqnarray}
where $R_{\mu\nu}$ is the Ricci tensor for the spacetime metric $g_{\mu\nu}$, $F_{\mu\nu}$ is
the Maxwell tensor, $\varphi$ is the dilaton field and $\alpha$ is the dilaton coupling parameter. For $\alpha=0$ (and $\varphi=0$)
we obtain the 5D Einstein-Maxwell equations.

We will  consider Kaluza-Klein spacetimes with  the symmetry group $R\times U(1)^2$ generated by the commuting Killing fields
$\xi$, $\zeta$  and $\eta$. Here $\xi$ is the asymptotically timelike Killing field and  $\zeta$  and $\eta$ are the axial Killing fields, respectively.
The Killing field $\eta$ will be associated with the fifth dimension. We also assume
that all the Killing fields  are mutually and  hypersurface orthogonal.
In this case, using adapted  coordinates in which $\xi=\partial/\partial {t}$, $\zeta=\partial/\partial \psi$ and $\eta=\partial/\partial \phi$
and canonical coordinate for the transverse  space the spacetime metric can be presented in the form

\begin{eqnarray}
ds^2 = g_{tt}dt^2 + g_{\rho\rho} (d\rho^2 + dz^2) + g_{\psi\psi}d\psi^2 + g_{\phi\phi}d\phi^2
\end{eqnarray}
where all the metric functions depend on the canonical coordinates $\rho$ and $z$ only.
We consider the case when  the electromagnetic and the dilaton fields respect the spacetime symmetries, i.e.
we want the electromagnetic gauge potential and the dilaton to have vanishing Lie derivatives along the Killing fields.
In particular for the electromagnetic field, we impose the following (local) ansatz
\begin{eqnarray}
A= A_{\phi}(\rho,z)d\phi
\end{eqnarray}
which is consistent with the spacetime symmetries.

After dimensional reduction along the Killing field $\eta$ we obtain an effective 4D theory whose "matter sector" (consisting of $A_{\phi}$ and  the norm of $\eta$) possesses group of hidden symmetries $GL(2,R)$ \cite{Y1}. Using the subgroup $SO(2) \subset GL(2,R)$
preserving the asymptotics we can generate exact solutions of the EMd-equations from exact solutions of the 5D vacuum Einstein equation.
For  a given  vacuum solution

\begin{eqnarray}\label{seedsolution}
ds_{E}^2= g^{E}_{tt}dt^2 + g^{E}_{\rho\rho} (d\rho^2 + dz^2) + g^{E}_{\psi\psi}d\psi^2 + g^{E}_{\phi\phi}d\phi^2
\end{eqnarray}
the $SO(2)$ transformation yields the following EMd-solution

\begin{eqnarray}\label{MAGSOLUTION}
&&ds^2 = \lambda^{1\over 1 + \alpha^2_{5}} \left[ g^{E}_{tt}dt^2 + g^{E}_{\rho\rho}(d\rho^2 + dz^2) + g^{E}_{\psi\psi} \psi^2\right]
+ \lambda^{-2\over 1 + \alpha^2_{5}} g^{E}_{\phi\phi} d\phi^2 ,\nonumber \\ \nonumber \\
&&e^{-{2(\varphi-\varphi_\infty)\over \sqrt{3}}}= \lambda^{\alpha_5\over 1+ \alpha^2_5 }, \\ \nonumber \\
&&A_{\phi}= e^{\alpha\varphi_{\infty}}{\sqrt{3}\tan\vartheta\over  2\sqrt{1+ \alpha^2_5}} {g^{E}_{\phi\phi}\over \lambda} + A^{0}_{\phi}
\nonumber
\end{eqnarray}
where
\begin{eqnarray}
\lambda= \cos^2\vartheta + g^{E}_{\phi\phi}\sin^2\vartheta,
\end{eqnarray}
$A^{0}_{\phi}$ and $\varphi_{\infty}$ are  constants and the $SO(2)$ -- parameter $\vartheta$ satisfies $-\pi/2 <\vartheta<\pi/2$. The parameter $\alpha_5$ is defined\footnote{In $D$-dimensional
spacetime we have $\alpha_{D}=\sqrt{{D-2\over {2(D-3)} }}\alpha$.} by $\alpha_5={\sqrt{3}\over 2}\alpha$.

Since the function $\lambda$ is positive everywhere the rod structure of the magnetized solution (\ref{MAGSOLUTION}) is inherited
from the seed solution (\ref{seedsolution}). We consider seed solutions with rod structure consisting of finite rods describing horizons and bubbles
and two semi-infinite rods describing the axes of  $\zeta=\partial/\partial\psi$. The Kaluza-Klein asymptotic condition and regularity conditions on the axes of
the Killing vectors  require\footnote{From now on the subscript or superscript "E" denotes quantities associated with the seed vacuum Einstein solutions. }

\begin{eqnarray}
(\Delta \psi)^{E}= 2\pi, \; \; \;
(\Delta \phi)^{E}=L_E,
\end{eqnarray}
to be satisfied for the seed solution. Then the regularity conditions for the magnetized solution are the following

\begin{eqnarray}
\Delta \psi= (\Delta \psi)^{E}=2\pi, \; \; \;
\Delta \phi = L= \left(\cos\vartheta\right)^{3/(1+ \alpha^2_5)} L_{E}.
\end{eqnarray}

Let us consider completely regular vacuum solution (\ref{seedsolution}) describing a configuration containing  at least one bubble.
Then the constant $A^{0}_{\phi}$ can not be arbitrary if we require the vector potential $A$ to be  regular everywhere.
Indeed on the bubble $g^{E}_{\phi\phi}=0$ and nonzero  $A^{0}_{\phi}$ means that the vector potential is divergent on the bubble.
In order for the vector potential to be globally well-defined we must impose $A^{0}_{\phi}=0$. This in turn means that the asymptotic
value of the vector potential is nonzero

\begin{eqnarray}
A_{\phi}^{\infty}=e^{\alpha\varphi_{\infty}} {\sqrt{3}\tan\vartheta\over 2\sqrt{1+ \alpha^2_5}}.
\end{eqnarray}

Let us further consider the asymptotic behaviour  of the solution.
For this purpose we introduce the asymptotic coordinates $r$ and $\theta$ defined as

\begin{eqnarray}\label{sphericalcoordinates}
\rho=r\sin\theta, \; \; \;
z=r\cos\theta.
\end{eqnarray}

After some algebra for the asymptotics  we find

\begin{eqnarray}
g_{tt} \approx -1 + {c_{t}\over r} , \;\; \;  g_{\phi\phi} \approx 1 + {c_\phi\over r}, \; \; \varphi \approx \varphi_{\infty} - {{\cal D}\over r}, \; \; A_{\phi} \approx A_{\phi}^\infty + {c_A\over r}
\end{eqnarray}
where

\begin{eqnarray}
&&c_t = c^{E}_t - {\sin^2\vartheta\over 1 + \alpha^2_{5}} c^{E}_{\phi}, \; \; \;\;\;\;\;\;\;
c_{\phi}= c^{E}_{\phi} - {2\sin^2\vartheta\over 1 + \alpha^2_{5}} c^{E}_{\phi}, \nonumber \\
&&{\cal D}=  {\sqrt{3}\alpha_{5}\sin^2\vartheta\over 2(1 + \alpha^2_5)} c^{E}_{\phi}, \; \; \; \; \;\;
c_A= e^{\alpha\varphi_{\infty}}{\sqrt{3}\sin\vartheta \cos\vartheta  \over 2\sqrt{1+ \alpha^2_5}} c^{E}_{\phi}.
\end{eqnarray}

The quantity ${\cal D}$ is the  dilaton (scalar) charge. An interpretation  of the quantity $c_A$ will be given below.

From the asymptotic behaviour we can compute the mass and the tension

\begin{eqnarray}
&&M={1\over 4}L(2c_t- c_\phi)= {1\over 4}L(2c^{E}_t- c^{E}_\phi)= {L\over L_E} M^{E} , \\
&&{\cal T} =  {1\over 4} (c_t - 2c_{\phi})= {1\over 4}[(c^{E}_t - 2c^{E}_{\phi}) + 3 {\sin^2\vartheta\over 1 + \alpha^2_5}c^{E}_{\phi}]=
{\cal T}^{E} + {3\over 4}{\sin^2\vartheta\over 1 + \alpha^2_5}c^{E}_{\phi} \\
&& ={\cal T}^{E} + {\sin^2\vartheta \over 1 + \alpha^2_5 }\left({M^{E}\over L_E} - 2{\cal T}^{E}\right),
\end{eqnarray}
where $M^{E}$ and ${\cal T}^{E}$ are the mass and the tension for the seed vacuum solution.

\section{Smarr-like relations for the mass and tension}

The mass and the tension can be also computed via the generalized Kommar integrals first introduced in \cite{Townsend:2001rg}.
These integrals are very useful in deriving Smarr-like relations. Here we present the generalized Kommar integrals
in a  form different but equivalent  to the original formulation in \cite{Townsend:2001rg}, namely

\begin{eqnarray}
M = - {L\over 16\pi} \int_{S^{2}_{\infty}} \left[2\star (\eta\wedge d\xi) - \star (\xi\wedge d\eta) \right]=
 - {L\over 16\pi} \int_{S^{2}_{\infty}} \left[2i_\eta \star d\xi - i_\xi \star d\eta \right],
\end{eqnarray}

\begin{eqnarray}
{\cal T}= - {1\over 16\pi} \int_{S^{2}_{\infty}} \left[\star (\eta\wedge d\xi) - 2\star (\xi\wedge d\eta) \right]=
 - {1\over 16\pi} \int_{S^{2}_{\infty}} \left[i_\eta \star d\xi - 2i_\xi \star d\eta \right],
\end{eqnarray}
where $\star$  is the Hodge dual and $i_X$ is the interior product of the vector field $X$ with an arbitrary form.

The generalized Kommar integrals allow us to define the intrinsic mass of  each object in the configuration \cite{KY}.
The intrinsic mass of each black hole  is given by

\begin{eqnarray}\label{BHKOMMARMASS}
M^{{\cal H}}_{i}= - {L\over 16\pi} \int_{{\cal H}_{i}} \left[2i_\eta \star d\xi - i_\xi \star d\eta \right]
\end{eqnarray}
where ${\cal H}_i$ is the 2-dimensional surface which is an intersection of the i-th horizon with a constant  $t$ and $\phi$
hypersurface. Analogously the intrinsic mass of each bubble is

\begin{eqnarray}\label{BKOMMARMASS}
M^{{\cal B}}_{j}= - {L\over 16\pi} \int_{{\cal B}_{j}} \left[2i_\eta \star d\xi - i_\xi \star d\eta \right].
\end{eqnarray}

One can show that the intrinsic masses of the black holes and bubbles are given by
\begin{eqnarray}
M^{{\cal H}}_{i}= {1\over 2}L l^{{\cal H}}_i, \; \; \;
M^{{\cal B}}_{j}= {1\over 4} L l^{{\cal B}}_j,
\end{eqnarray}
where $l^{{\cal H}}_i$ and $l^{{\cal B}}_j$ are the lengths of the horizon and bubble rods, respectively. It was also  shown in \cite{KY} that

\begin{eqnarray}
M^{{\cal H}}_{i}= {1\over 4\pi} \kappa_{{\cal H}_i} {\cal A}_{{\cal H}_i}, \;\;\;
M^{\cal B}_{j}= {L\over 8\pi}\kappa_{{\cal B}_j} {\cal A}_{{\cal B}_j},
\end{eqnarray}
where $\kappa_{{\cal H}_i}$ and ${\cal A}_{{\cal H}_i}$ are the surface gravity and the area of the i-th horizon and the surface gravity and area of j-th bubble. The surface gravity and the area for a bubble were first introduced  in \cite{Kastor:2008wd}. The bubble surface gravity is defined by

\begin{eqnarray}
\kappa^2_{\cal B}= {1\over 2} \nabla_{[\mu}\eta_{\nu]} \nabla^{[\mu}\eta^{\nu]}
\end{eqnarray}
where the right hand side is evaluated on the bubble. The reader might consult \cite{Kastor:2008wd} for other equivalent definitions. The bubble area is given by

\begin{eqnarray}
{\cal A}_{\cal B}= \int_{{\cal B}}\sqrt{|g_{tt}| g_{\rho\rho}g_{\psi\psi}}\,dzd\psi.
\end{eqnarray}

For regular (smooth ) bubbles (i.e. bubbles without conical singularities), the case we consider here, one can show that

\begin{eqnarray}
\kappa_{\cal B}= {2\pi\over  L}.
\end{eqnarray}
The bubble area for our solution can be easily calculated and the result is

\begin{eqnarray}
{\cal A}_{\cal B}={\cal A}^{E}_{\cal B} {L\over L_E}.
\end{eqnarray}

Concerning the area and the surface gravity of the horizons we find

\begin{eqnarray}
\kappa_{{\cal H}_i}= \kappa_{{\cal H}_i}^{E}, \;\;\;\;\;  {\cal A}_{{\cal H}_i}={\cal A}_{{\cal H}_i}^{E} {L\over L_E}.
\end{eqnarray}

Let us note that although the $SO(2)$-transformations preserve the determinant of the horizon metric, the horizon area
is different from that of the seed solution since the period of $\phi$ for our solution is different from the period of the seed solution.

We proceed further with deriving Smarr-like relations for the tension and the mass.
Using the Stokes theorem we find that the tension can be written as a bulk  integral over a constant $t$ and $\phi$ hypersurface $\Sigma$
and surface integrals over black hole horizons and bubbles

\begin{eqnarray}
{\cal T} L &=& -{L\over 16\pi} \sum_i\int_{{\cal H}_i} \left(i_\eta \star d\xi - 2i_\xi\star d\eta\right)
-{L\over 16\pi} \sum_j\int_{{\cal B}_j} \left(i_\eta \star d\xi - 2i_\xi\star d\eta\right) \\
&&-{L\over 16\pi} \int_{\Sigma}d\left(i_\eta \star d\xi - 2i_\xi\star d\eta \right) \nonumber
\end{eqnarray}
where we have taken into account that $\partial\Sigma =S^{2}_{\infty} -\sum_i {\cal H}_i - \sum_j {\cal B}_j$. Using the definitions
(\ref{BHKOMMARMASS}) and (\ref{BKOMMARMASS}), the Killing symmetries and the identity $d\star d\xi=2\star R[\xi]$ for an arbitrary
Killing field, we have

\begin{eqnarray}
{\cal T} L = {1\over 2} \sum_i M^{{\cal H}}_i + 2\sum_j M^{{\cal B}}_j + {L\over 8\pi} \int_{\Sigma} \left(i_{\eta}\star R[\xi] - 2i_\xi \star R[\eta] \right)
\end{eqnarray}
where $R[X]$ is Ricci 1-form with respect to the vector field $X$. Making advantage of the field equations (\ref{EMDFE}) we obtain

\begin{eqnarray}
\star R[\xi] = - 2e^{-2\alpha\varphi} \left( -{2\over 3}i_{\xi}F\wedge \star F + {1\over 3} F\wedge i_{\xi}\star F \right)
\end{eqnarray}
and the same expression for $\star R[\eta]$, however with $\xi$ replaced by $\eta$. Hence we find

\begin{eqnarray}
i_{\eta}\star R[\xi] - 2i_\xi \star R[\eta]= 2e^{-2\alpha\varphi} i_\eta F \wedge i_\xi\star F
\end{eqnarray}
and therefore

\begin{eqnarray}
{\cal T} L = {1\over 2} \sum_i M^{{\cal H}}_i + 2\sum_j M^{{\cal B}}_j + {L\over 4\pi} \int_{\Sigma} e^{-2\alpha\varphi} i_\eta F \wedge i_\xi\star F.
\end{eqnarray}

Since $A$ is globally well-defined we can write $F=dA$.  Taking further into account  that ${\cal L}_{\eta}A=0$ and $d\left(e^{-2\alpha\varphi}i_{\xi}\star F\right)=0$ as a consequence of the field equations, we have

\begin{eqnarray}
\int_{\Sigma} e^{-2\alpha\varphi} i_\eta F \wedge i_\xi\star F= -  \int_{\Sigma} d\left(i_{\eta}A e^{-2\alpha\varphi}i_\xi \star F \right)= - \int_{S^2_{\infty}} (i_{\eta}A) e^{-2\alpha\varphi}i_\xi \star F
\\ + \sum_i\int_{{\cal H}_{i}} (i_{\eta}A) e^{-2\alpha\varphi}i_\xi \star F
+ \sum_j \int_{{\cal B}_{j}} (i_{\eta}A) e^{-2\alpha\varphi}i_\xi \star F .\nonumber
\end{eqnarray}

Using the invariance under the Killing field $\zeta$ we have

\begin{eqnarray}
\int_{\Sigma} e^{-2\alpha\varphi} i_\eta F \wedge i_\xi\star F= -  \int_{\Sigma} d\left(i_{\eta}A e^{-2\alpha\varphi}i_\xi \star F \right)= - 2\pi \int_{S^2_{\infty}/U(1)} (i_{\eta}A) e^{-2\alpha\varphi}i_{\zeta}i_\xi \star F
\\ + 2\pi\sum_i\int_{{\cal H}_{i}/U(1)} (i_{\eta}A) e^{-2\alpha\varphi}i_{\zeta}i_\xi \star F
+ 2\pi\sum_j \int_{{\cal B}_{j}/U(1)} (i_{\eta}A) e^{-2\alpha\varphi}i_{\zeta}i_\xi \star F . \nonumber
\end{eqnarray}

On the bubbles $\eta=0$ which shows that the third term on the right hand side vanishes.
For simplicity we shall consider black holes with bifurcate Killing  horizons. Therefore  the second term also vanishes and we obtain

\begin{eqnarray}
{\cal T} L = {1\over 2} \sum_i M^{{\cal H}}_i + 2\sum_j M^{{\cal B}}_j
 - {L\over 2}A_{\phi}^{\infty}  \int_{S^2_{\infty}/U(1)} e^{-2\alpha\varphi}i_{\zeta}i_\xi \star F .
\end{eqnarray}

Using the asymptotic behaviour  we find

\begin{eqnarray}
i_{\zeta}i_{\xi}\star F|_{S^2_{\infty}}=-c_A\sin\theta d\theta
\end{eqnarray}
and therefore

\begin{eqnarray}
 \int_{S^2_{\infty}/U(1)}  e^{-2\alpha\varphi} i_{\zeta}i_\xi \star F =-e^{-2\alpha\varphi_{\infty}} \int^{\pi}_{0} c_A\sin\theta d\theta=-2e^{-2\alpha\varphi_{\infty}}c_A.
\end{eqnarray}

In this way we obtain

\begin{eqnarray}
{\cal T} L = {1\over 2} \sum_i M^{{\cal H}}_i + 2\sum_j M^{{\cal B}}_j + L A_{\phi}^{\infty}e^{-2\alpha\varphi_{\infty}}c_A.
\end{eqnarray}

In order to interpret the last term in the above formula we shall proceed as follows. Let us consider the
right-most (or the left-most) bubble rod which we denote by  $[z_1,z_2]$.
Then we fix a point $z_{f}>>z_2$ lying on the rod corresponding to the axis of $\zeta$ and consider the path $\hat \gamma= [z_2, z_{f}]$.
We lift $\hat \gamma$ to a path $\gamma$ in the spacetime manifold and define $C$ to be the 2-dimensional surface obtained from  $\gamma$
by acting with the isometries generated by $\eta$. Since $\eta|_{z_2}=0$ the  2-surface $C$ has the topology of a disk.
In this way we can define the magnetic flux through the 2-surface $C$

\begin{eqnarray}
\Psi_C = \int_{C} F = \int_{\partial C} A= LA_{\phi}(z_f).
\end{eqnarray}

Therefore the quantity

\begin{eqnarray}\label{flux}
\Psi = LA_{\phi}^{\infty}=\lim_{z_f\to \infty} \Psi_C
\end{eqnarray}
can be interpreted as the magnetic flux through  the infinitely extended 2-surface $C$. Respectively, the integral

\begin{eqnarray}
J =-{1\over 4\pi}\int_{S^2_{\infty}} e^{-2\alpha\varphi} i_{\xi}\star F=e^{-2\alpha\varphi_{\infty}}c_A
\end{eqnarray}
should be interpreted as an effective current sourcing  of the magnetic field.
The Smarr-like relation can then be rewritten in the following form

\begin{eqnarray}\label{Smarr-Tension}
{\cal T} L = {1\over 2} \sum_i M^{{\cal H}}_i + 2\sum_j M^{{\cal B}}_j + J\Psi.
\end{eqnarray}

Following the same method as for the tension one can show that

\begin{eqnarray} M= \sum_i M^{{\cal H}} + \sum_j M^{{\cal B}}_j - {L\over 4\pi} \int_{\Sigma} e^{-2\alpha\varphi}i_\xi F \wedge i_\eta \star F .
\end{eqnarray}

In the case under consideration $i_\xi F=0$ and  for the ADM mass  we find

\begin{eqnarray}\label{Smarr-MASS}
M= \sum_i M^{{\cal H}}_{i} + \sum_j M^{{\cal B}}_j.
\end{eqnarray}

\section{Mass and tension first law}

Our next goal is to derive the first law for the black hole-bubbles configurations under consideration.
In our derivation we shall follow the Wald's approach \cite{Wald} making use of the Noether current.

Our diffeomorphism covariant theory is derived from  the  Lagrangian
\begin{eqnarray}
 {\mathbf  L} = \star R - 2d\varphi \wedge \star d\phi - 2e^{-2\alpha\varphi} F\wedge \star F.
\end{eqnarray}
When the field equations are satisfied the first order variation of the Lagrangian is given by

\begin{eqnarray}
\delta {\mathbf L} = d {\Theta}
\end{eqnarray}
where

\begin{eqnarray}
\Theta = \star \,\upsilon - 4 (\star d\varphi)\delta \varphi + 4\left(e^{-2\alpha\varphi} \star F\right) \wedge \delta A
\end{eqnarray}
and

\begin{eqnarray}
\upsilon_\mu = \nabla^{\nu}\delta g_{\mu\nu} - g^{\alpha\beta}\nabla_{\mu}\delta g_{\alpha\beta}.
\end{eqnarray}

The Noether current ${\cal I}^{X}$ associated  with a diffeomorphism generated by an arbitrary smooth vector field $X$ is

\begin{eqnarray}
{\cal I}^{X}= \Theta(\Gamma, {\cal L}_{X}\Gamma) - i_X {\mathbf L},
\end{eqnarray}
where the fields $g_{\mu\nu}, A_{\mu}, \varphi$ are collectively denoted by $\Gamma$. The current ${\cal I}^{X}$ satisfies
$d{\cal I}^{X}=0$ when the field equations are satisfied. Since ${\cal I}^{X}$ is closed there exists a 3-form ${\cal N}^{X}$ (Noether charge 3-form)
such that ${\cal I}=d{\cal N}^{X}$.

Now, let $\Gamma$ is a solution to the field equations (\ref{EMDFE}) and let $\delta \Gamma$ is a
linearized perturbation satisfying  the linearized equations of the Einstein-Maxwell-dilaton gravity.
For simplicity we will also assume that ${\cal L}_{\xi}\delta \Gamma={\cal L}_{\eta}\delta \Gamma=0$.  Then, choosing $X$ to be a Killing field one can show that \cite{Wald}

\begin{eqnarray}
\delta d{\cal N}^{X}=di_X\Theta.
\end{eqnarray}

In the case under consideration we need the Noether forms ${\cal N}^{\xi}$ and ${\cal N}^{\eta}$. After some calculations it can be shown that
they are given by

\begin{eqnarray}
&&{\cal N}^{\xi}= - \star d\xi, \\  \nonumber \\
&&{\cal N}^{\eta}= - \star d\eta - 4i_\eta A \left(e^{-2\alpha\varphi}\star F \right).
\end{eqnarray}

In fact what we need are the 2-forms $i_\eta {\cal N}^{\xi}$  and $i_\xi {\cal N}^{\eta}$. For them one can show that

\begin{eqnarray}\label{Noetheridentities}
\delta \left( di_{\eta}{\cal N}^{\xi}\right)=di_{\eta}i_{\xi}\Theta, \;\;\; \; \; \delta \left(di_{\xi}{\cal N}^{\eta}\right)= - di_{\eta}i_{\xi}\Theta.
\end{eqnarray}

It turns out useful to combine (\ref{Noetheridentities}) to a single equality

\begin{eqnarray}\label{combinedidentity}
 \delta  \left(2 di_{\eta}{\cal N}^{\xi} - di_{\xi}{\cal N}^{\eta} \right)= 3di_{\eta}i_{\xi}\Theta . \end{eqnarray}

Integrating on $\Sigma$ and using the Stokes  theorem we obtain

\begin{eqnarray}\label{combinedidentity1}
\delta \int_{\partial \Sigma} \left(2 i_{\eta}{\cal N}^{\xi} - i_{\xi}{\cal N}^{\eta} \right)=
3\int_{\partial \Sigma} di_{\eta}i_{\xi}\Theta
\end{eqnarray}
 where $\partial \Sigma=S^{2}_{\infty} - \sum_i {\cal H}_i - \sum_j {\cal B}_j$.   Taking into account
the explicit form of the Noether forms  we find

\begin{eqnarray}
&&\int_{S^2_{\infty}} \left(2 i_{\eta}{\cal N}^{\xi} - i_{\xi}{\cal N}^{\eta} \right)= 16\pi {M\over L} -16\pi A^{\infty}_{\phi} J, \\
&&\int_{{\cal H}_i} \left(2 i_{\eta}{\cal N}^{\xi} - i_{\xi}{\cal N}^{\eta} \right)=  16\pi {M_{{\cal H}_i}\over L} , \\
&& \int_{{\cal B}_j} \left(2 i_{\eta}{\cal N}^{\xi} - i_{\xi}{\cal N}^{\eta} \right)= 16\pi {M_{{\cal B}_j}\over L}. \end{eqnarray}
Respectively, for  $i_\eta i_\xi \Theta$ we have

\begin{eqnarray}
&&\int_{S^2_{\infty}}  i_\eta i_\xi \Theta = -4\pi (\delta c_t - \delta c_\phi) + 16\pi {\cal D} \delta\varphi_{\infty} - 16\pi J\delta A^{\infty}_{\phi}, \\
&&\int_{{\cal H}_i}  i_\eta i_\xi \Theta = 2 {{\cal A}_{H_i}\over L }\delta \kappa_{H_i}, \\
&& \int_{{\cal B}_j}i_\eta i_\xi \Theta = 2{{\cal A}_{B_j}\over L } \delta \kappa_{{\cal B}_j},
\end{eqnarray}
where we have taken into account that the dilaton charge can be expressed in the form

\begin{eqnarray}
{\cal D}= {1\over 4\pi} \int_{S^2_{\infty}} i_\eta i_\xi\star d\phi.
\end{eqnarray}

Substituting  these results in (\ref{combinedidentity1})   and using the relation (\ref{Smarr-MASS})   we obtain

\begin{eqnarray}
 {3\over 4} (\delta c_t - \delta c_\phi)= 3{\cal D}\delta\varphi_{\infty} - 2J\delta A^{\infty}_{\phi} + A^{\infty}_{\phi}\delta J
- {3\over 8\pi} \sum_i {{\cal A}_{{\cal H}_i}\over L}\delta \kappa_{{\cal H}_i}
- {3\over 8\pi} \sum_j {{\cal A}_{{\cal B}_j}\over L}\delta \kappa_{{\cal B}_j}.
\end{eqnarray}

The next step is to take into account that $3/4(\delta c_t - \delta c_\phi)=\delta (M/L) + \delta {\cal T}$ and
to express $\delta {\cal T}$ from the Smarr-like relation (\ref{Smarr-Tension}) which gives

\begin{eqnarray}
\delta \left({M\over L}\right) &=&  - {1\over 2\pi} \sum_{i} {{\cal A}_{{\cal H}_i}\over L } \delta \kappa_{{\cal H}_i}
- {1\over 8\pi}\sum_i \kappa_{{\cal H}_i} \delta  \left({{\cal A}_{{\cal H}_i}\over L }\right)  -
 {5\over 8\pi} \sum_{j} {\cal A}_{{\cal B}_j} \delta \kappa_{{\cal B}_j} \\
&&- {1\over 4\pi}\sum_j \kappa_{{\cal B}_j} \delta {\cal A}_{{\cal B}_j}  - 3J \delta A^{\infty}_{\phi}
+ 3{\cal D} \delta\varphi_{\infty}. \nonumber
\end{eqnarray}

Now using the Smarr-like relation (\ref{Smarr-MASS}) we also find

\begin{eqnarray}
\delta \left({M\over L}\right) &=&   {1\over 4\pi} \sum_{i} {{\cal A}_{{\cal H}_i}\over L } \delta \kappa_{{\cal H}_i}
+ {1\over 4\pi}\sum_i \kappa_{{\cal H}_i} \delta  \left({{\cal A}_{{\cal H}_i}\over L }\right)  +
 {1\over 8\pi} \sum_{j} {\cal A}_{{\cal B}_j} \delta \kappa_{{\cal B}_j} \\
&&+ {1\over 8\pi}\sum_j \kappa_{{\cal B}_j} \delta {\cal A}_{{\cal B}_j} . \nonumber
\end{eqnarray}

Combining the above equalities we obtain

\begin{eqnarray}
\delta \left( {M\over L}  \right) = \sum_i {\kappa_{{\cal H}_i}\over 8\pi} \delta \left({{\cal A}_{{\cal H}_{i}}\over L}\right)
- {1\over 8\pi}\sum_j {\cal A}_{{\cal B}_j} \delta \kappa_{{\cal B}_j} - J\delta A^{\infty}_{\phi} + {\cal D}\delta\varphi_{\infty}
\end{eqnarray}
which in view of the relations (\ref{flux}) and (\ref{Smarr-Tension}) gives

\begin{eqnarray}\label{MASSFIRSTLAW}
\delta M = \sum_i {\kappa_{{\cal H}_i}\over 8\pi} \delta {\cal A}_{{\cal H}_{i}}
- {1\over 8\pi} \sum_j {\cal A}_{{\cal B}_{j}}\delta (\kappa_{{\cal B}_j}L)  - J \delta \Psi + L{\cal D}\delta\varphi_{\infty} + {\cal T}\delta L.
\end{eqnarray}

This is the desired form of the mass first law. Once having the mass first law, the tension first law can be easily found and the result is

\begin{eqnarray}  \delta {\cal T}= {1\over 8\pi} \sum_j \kappa_{{\cal B}_j}\delta {\cal A}_{{\cal B}_j}
- {1\over 8\pi}\sum_j{{\cal A}_{{\cal H}_i}\over  L} \delta \kappa_{{\cal H}_i}    + {\Psi\over L}\delta J + {\cal D}\delta \varphi_{\infty}.
\end{eqnarray}

Let us note that in the limit of vanishing magnetic field our formulas for the mass and the tension first
law are reduced to the mass and tension first law  for vacuum black hole-bubble configurations derived by other means in \cite{Kastor:2008wd}.
We also note that for regular bubbles the second term on the right hand side of (\ref{MASSFIRSTLAW})
does not give any contribution since $\kappa_{{\cal B}_j}L=2\pi$.

So far our considerations were general without specifying the explicit configuration of black holes and Kaluza-Klein bubbles.
It is easy  specific example to be constructed. As for example, one may take the exact solution describing
a sequence of black rings and Kaluza-Klein bubbles \cite{EHO} which substituted in the general formulae (\ref{MAGSOLUTION}) will produce a magnetized configuration of sequenced black rings and bubbles. Here, as an explicit example,  we will present
the simplest magnetized configuration consisting of a single bubble which in the standard spherical coordinates (\ref{sphericalcoordinates}) is given by

\begin{eqnarray}
&&ds^2= \left(1 - {L_{E}\sin^2\vartheta\over 4\pi r}\right)^{1\over 1+ \alpha^2_5}\left[-dt^2
+ {dr^2\over 1 - {L_{E}\over 4\pi r} } + r^2d\Omega^2_2\right]  \nonumber \\ && + \left(1 - {L_{E}\sin^2\vartheta\over 4\pi r}\right)^{-2\over 1+ \alpha^2_5} \left(1 - {L_{E}\over 4\pi r}\right) d\phi^2, \\ \nonumber \\
&&e^{-2(\varphi - \varphi_\infty)\over \sqrt{3}} = \left(1 - {L_{E}\sin^2\vartheta\over 4\pi r}\right)^{\alpha_5\over 1+ \alpha^2_5}, \\ \nonumber \\
&&A_{\phi}= e^{\alpha\varphi_{\infty}}{\sqrt{3}\tan\vartheta \over 2\sqrt{1+ \alpha^2_5}} {r - {L_E\over 4\pi}\over r- {L_E\sin^2\vartheta\over 4\pi}}.
\end{eqnarray}

\section{Conclusion}

In this work we presented a general class of exact solutions to the 5D equations of Einstein-Maxwell-dilaton gravity describing
configurations with black holes and Kaluza-Klein bubbles magnetized along the compact dimension. We also derived the Smarr-like
relations and mass and tension first law for such configurations. These results can be straightforwardly generalized to
Kaluza-Klein spacetimes with number of dimensions $D>5$.

The solutions obtained here are not completely characterized by the interval (rod) structure. Since the electric and magnetic  charges are
zero by construction  they can not be employed to classify the solutions. Without rigorous proof, which will be given elsewhere,
but as it is clear, we note that  the solutions can be classified by the rod structure and magnetic flux $\Psi$. In this way, in the generalization of
the  uniqueness theorem of \cite{HY3} to the case with Maxwell fields a new data should be included in addition to the rod structure, angular
momenta and charges -- the magnetic fluxes.

We shall finish with some prospects  for future work. It will be  interesting to consider configurations with (magnetically charged) Kaluza-Klein
bubbles and
magnetically charged (dipole) black rings or black strings. We expect that the interaction between the magnetic field and the charges
yields new regular balanced solutions with new interesting properties. Some results in these directions have already been obtained \cite{YN}  and they will be
published elsewhere.

\section*{Acknowledgements}
 The partial support by the
Bulgarian National Science Fund under Grants  VUF-201/06 and DO 02-257, is  acknowledged.

\end{document}